# Translation and rotation of a spherical particle in a turbulent boundary layer

Yi Hui Tee[1*], Diogo Barros[1], Ellen K. Longmire[1]
1: Aerospace Engineering and Mechanics, University of Minnesota, Minneapolis, MN 55455, USA
* Correspondent author: teexx010@umn.edu

ABSTRACT

Three-dimensional particle tracking experiments were conducted in a turbulent boundary layer with friction Reynolds number $Re_\tau$ of 700 and 1300. Two finite size spheres with specific gravities of 1.003 (P1) and 1.050 (P2) and diameters of 60 and 120 wall units were released individually from rest on a smooth wall. The spheres were marked with dots all over the surface to monitor their translation and rotation via high-speed stereoscopic imaging. The 3D particle translational and rotational dynamics were examined. The spheres accelerated strongly after release over streamwise distances of one boundary layer thickness before approaching an approximate terminal velocity. Initially, sphere P1, which had Reynolds numbers $Re_p$ of 800 and 1900, always lifts off from the wall. Similar behavior was observed occasionally for sphere P2 with initial $Re_p$ of 1900. The spheres that lifted off reached an initial peak in height before descending towards the wall. The sphere trajectories exhibited multiple behaviors including saltation, resuspension (either near the wall or farther from the wall) and sliding motion with small random bouncing depending on both $Re_\tau$ and specific gravity. The lighter sphere at $Re_\tau = 1300$, which remained suspended above the wall during most of its trajectory, propagated with the fastest streamwise velocity. By contrast, the denser sphere at $Re_\tau = 700$, which mostly slid along the wall, propagated with the slowest streamwise velocity. After the spheres approached an approximate terminal velocity, many experienced additional lift-off events that were hypothesized to be driven by hairpins or coherent flow structures. Spheres were observed to rotate about all three coordinate axes. While the mean shear may induce a rotation about the spanwise axis, near-wall coherent structures and the sphere's wake might drive the streamwise and wall-normal rotations. In all cases where the sphere propagates along the wall, sliding motion, rather than forward rolling motion, is dominant.

## 1. Introduction

Particle-laden turbulent flows occur widely in environmental and industrial processes. In wall-bounded flows, fluid shear and pressure forces lift particles off the wall and drag them along it. While propagating with the fluid, coherent structures in the boundary layer can have significant effects on particle suspension, deposition and transport. For particles of finite size, wall interactions are also significant. Depending on the surrounding conditions, particles can either slide or roll along the bounding surface and lift-off from or collide with the wall. Here, friction and restitution coefficients play important roles in the particle dynamics. Therefore, the presence of both particle-turbulence and particle-wall interactions will significantly affect the translation and rotation of particles in wall-bounded turbulent flows.

Early investigations conducted by Francis (1973) reported on the existence of three motion modes for single heavy grains transported over a planar rough bed. Grains of multiple geometries including spherical particles might exhibit rolling, saltation or suspension behavior. In all modes, spinning motion was observed, and maximum angular velocities were visualized close to particle lift-off. White and Schulz (1977) studied the motion of spherical glass microbeads in air using high-speed movies. They suggested that spinning-associated Magnus forces could be a non-negligible part of the lift force exerted on the particles. However, no angular velocity measurements were conducted by those authors. Sumer and Oguz (1978) described the motion of individual slightly "heavy" particles above a smooth wall in a turbulent channel flow. In particular, they attempted to explain how the flow structures in the near-wall region prevented suspended particles from settling towards the wall. They observed that the particles alternately moved upwards and downwards, and this phenomenology was supported by the bursting low-speed streak model proposed by Offen and Kline (1975). Later, Sumer and Deigaard (1981) further characterized the motion of particles with densities 0.2 - 3% higher than the carrying fluid for particle diameters of $d^+$~ 50. They concluded that the motion of suspended particles was similar over rough and smooth walls.

More recent experiments in this area have included those of Kaftori *et al.* (1995), Niño and Garcia (1996) and van Hout (2013) who investigated the resuspension of particles in turbulent wall-bounded flows using various flow visualization techniques. These studies concluded that the particle resuspension and deposition events in the near wall region were strongly influenced by coherent flow structures. From the time-resolved PIV and PTV experiments by van Hout (2013), it was observed that polystyrene beads of $d^+ \approx 10$, in all cases, lifted off the wall due to positive shear induced by ejection events generated by passing vortex cores. Once lifted beyond the viscous sublayer, the particle either stayed suspended in the fluid or saltated along the wall depending on the type of coherent structures that it encountered. In general, these previous experiments did not isolate individual particle motion from possible effects due to particle-particle interactions.

A key factor in particle resuspension is the lift force acting on a particle. Saffman (1965)'s lift force expression provides a good approximation for a sphere in a fluid away from walls at low particle Reynolds number. However, lift forces acting on particles in a wall-bounded turbulent flow at finite Reynolds number are poorly understood. In this context, Hall (1988), Mollinger and Nieuwstadt (1996), and Schmeeckle *et al.* (2007), among others, examined experimentally the fluid forces acting on a stationary particle lying on the wall in a turbulent boundary layer. Hall (1988) measured the lift force using a force transducer. The lift data showed that for $d^+$ between 3.6 and 140 wall units and particle Reynolds number, $Re_p$ between 6.5 and 1250, the normalized mean lift force could be approximated by $F^+ = (20.90 \pm 1.57)(d^+/2)^{2.31\pm0.02}$. On the other hand, simulations of finite-size particles in turbulent boundary layers are very limited. For instance, Zeng *et al.* (2008) performed a fully resolved direct numerical simulation (DNS) to compute the fluid forces acting on a fixed particle located close to the wall in a turbulent channel flow. For $d^+ < 100$ and a friction Reynolds number of $Re_\tau = 178.12$, the lift contribution became significant as the stationary particle approached the wall in the viscous sublayer region. Also, due to the presence of the wall, the computed drag force was slightly higher than the standard drag correlation.

Tracking the dynamics of discrete particles with significant size is fundamental to characterize fluid-solid interactions occurring in those flows. Experimentally, determining both the translation and rotation of a given particle within a 3D domain with accuracy poses a challenging problem. In the specific case of finite-size spherical particles, this problem has been addressed previously using multiple methods. Printing specific patterns over the surface of a sphere allows one to obtain its absolute orientation by comparing images acquired with high-speed cameras to synthetic projections. Information from two cameras positioned perpendicularly was coupled to reconstruct the three-dimensional location of the sphere by Zimmermann *et al.* (2011) and Mathai *et al.* (2016). An alternative method to track the motion of large particles has been to introduce visible tracers into the interior of transparent spheres. Klein *et al.* (2013) used fluorescent markers embedded in a hydrogel sphere to calculate its translation and angular velocity. To track the location of each tracer, 3D Lagrangian tracking was employed with a three-camera arrangement: the center of the sphere was computed using the equation of a sphere by the knowledge of its radius and at least four markers. The relative coordinates of the tracers (markers) with respect to the center allow one to compute the optimal rotation matrix by applying an optimization algorithm [Kabsch (1976, 1978)]. Finally, the Euler rotation rates could be extracted from the optimal rotation matrix to determine the angular velocity of the sphere.

In the present study, we conduct three-dimensional particle tracking experiments in a water channel to investigate the motion of single, finite-size spheres released in a turbulent boundary layer along a smooth wall. We study particle-eddy and particle-wall interactions by varying both boundary layer friction velocity and particle specific gravity. In particular, the 3D trajectories of the particle and its rotational dynamics are investigated under different conditions. The dynamics of particle lift-off and motion are discussed. To track the motion of the spheres, we employ the methodology proposed recently by Barros *et al.* (2018) where non-transparent spheres are considered. The displacement of the sphere and its angular velocity are computed using 3D Lagrangian tracking of markers painted arbitrarily over the geometry.

## 2. Methodology
### 2.1 Experimental set up

The experiments were conducted in a closed-return water channel facility at the University of Minnesota. The test section of the channel, which is constructed of glass, is 8 m long x 1.12 m wide. A trip-wire of 3 mm diameter was located at the entrance of the test section to trigger the development of a turbulent boundary layer along the bottom wall. Hereafter $x$, $y$ and $z$ define the streamwise, wall-normal, and spanwise directions, respectively.

To achieve a repeatable and controllable initial experiment condition, magnetic spheres molded from a mixture of wax and iron oxide were used. Iron oxide was added to make the particles magnetic and to control their densities. Multiple dots were marked all over the surface of the sphere to monitor its translation and rotation. The sphere's surface is black and opaque providing good contrast with the white dots produced with commercial oil-based pen. The markers were painted manually at arbitrary locations on the surface (see fig. **1**(a) and (b)). The sphere was then held in place on the bottom wall of the water channel by a cubic magnet located flush to the external wall of the channel. Deactivating the magnet allowed the sphere to move with the fluid along the channel. In all experiments, one sphere was released at a location 4.2 m downstream of the trip wire. This streamwise location will be considered as $x = 0$. A screen was located at the end of the test section to capture the sphere and prevent it from recirculating around the channel.

Two pairs of Phantom v210 cameras from Vision Research Inc. were arranged in stereoscopic configurations to monitor the three-dimensional particle trajectories over a relatively long field of view (see fig. **1**(c)). All cameras were fitted with 105mm Nikon Micro-Nikkor lenses

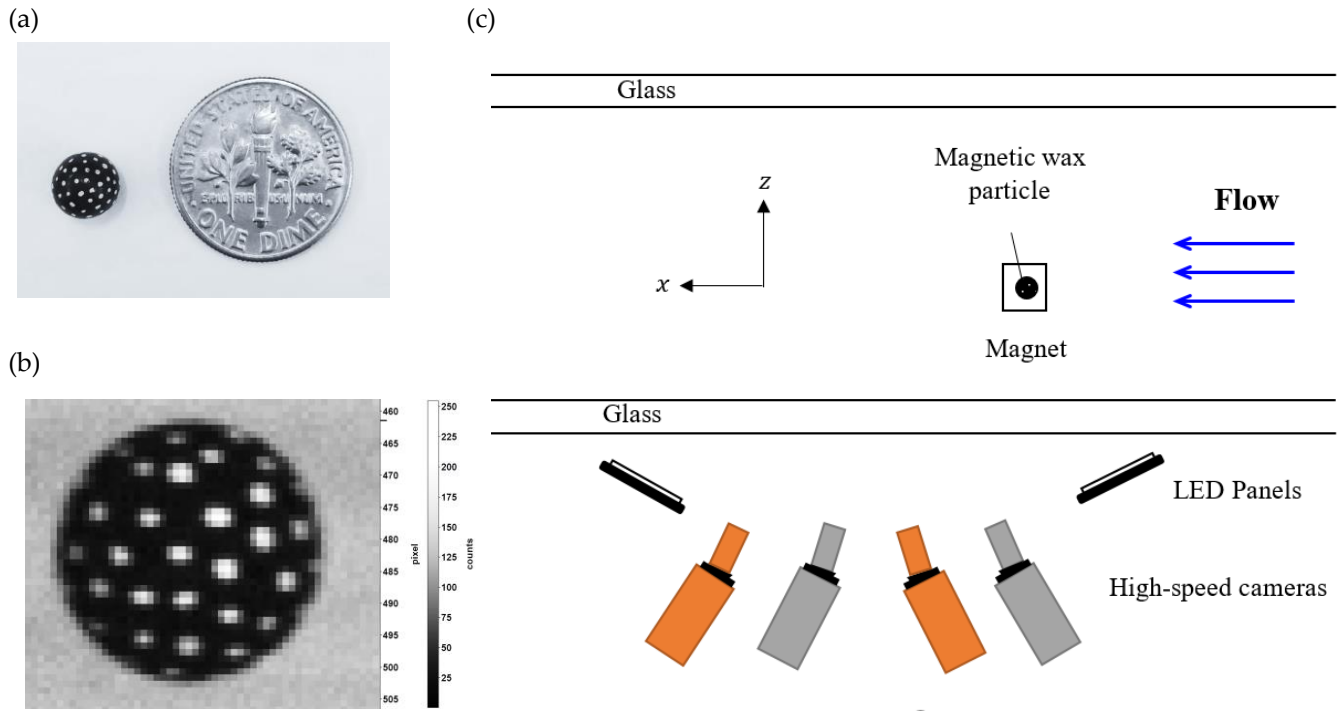

**Fig. 1** (a) Photograph of a 6.35 mm sphere painted with dots (left) when compared to a dime (right). (b) Pixelated sphere (grayscale between 0 to 256 counts) plotted with scale in pixels. The sphere diameter is 43 pixels with scale factor of 0.148 mm/px. (c) Top view of the experimental setup: two pairs of high-speed cameras were aligned in stereoscopic configuration for capturing the trajectory and rotation of a marked sphere over a long field of view.

with aperture f/16. Scheimpflug mounts were added to all cameras so that the images are uniformly focused across the fields of view. The experiments were captured at a sampling frequency of 480 Hz with image resolution of 1280 x 800 pixels. Three white LED panels illuminated the domain considered.

## 2.2 Experimental Parameters

We conducted experiments at two free-stream velocities, $U_\infty$ of 0.22 m/s and 0.49 m/s, which correspond to friction Reynolds numbers, $Re_\tau$ of 700 and 1300 at the particle release location. The mean flow statistics of the unperturbed turbulent boundary layers were determined from planar PIV measurements in streamwise wall-normal planes. Silver-coated hollow glass spheres of 13 $\mu m$ with density of 1.6 x $10^3$ kg/$m^3$ were used to seed the flow. A double-pulsed New Wave Solo II Nd:YAG 532 nm laser system provided the flow illumination. The laser beam was expanded through a series of spherical and cylindrical lenses to form a laser sheet of 1 mm thickness. A total of 2000 independent image pairs were acquired at a repetition rate of 1.8 Hz using a TSI Powerview Plus 4MP PIV camera. The images were processed in Davis 8.3 from LaVision using an interrogation window of 32 x 32 pixels with 50% overlap. The mean and root-mean-square streamwise velocity profiles at both $Re_\tau$ values are presented in fig. 2. The profiles agree well with the DNS simulations from Jiménez *et al.* (2010). At $Re_\tau$ of 700 and 1300, the boundary layer thickness, $\delta$ was 0.073 m and 0.066 m, respectively.

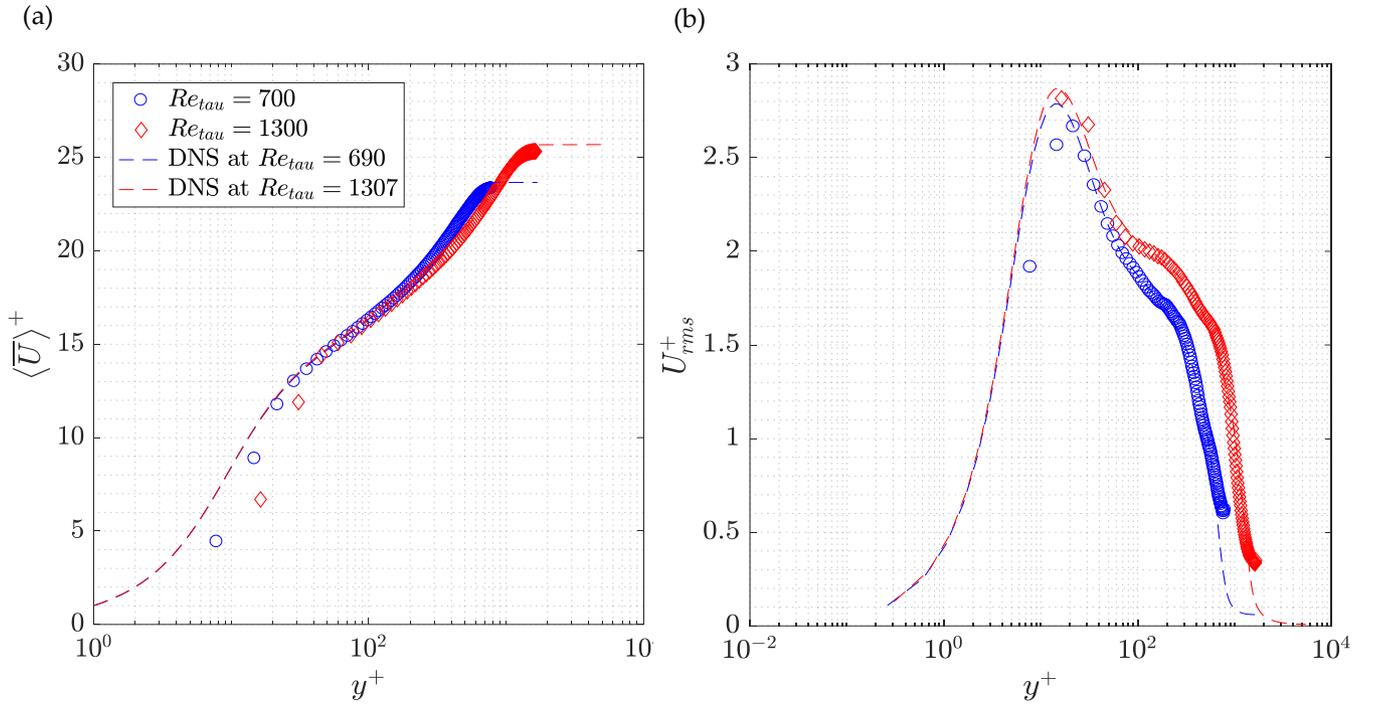

**Fig. 2** Statistics of the unperturbed turbulent boundary layers. (a) Mean and (b) root-mean-square streamwise velocity profiles at different $Re_\tau$ normalized by wall units. DNS profiles from Jiménez *et al.* (2010).

Two nearly neutrally buoyant spherical particles were considered to limit the effect of gravity on the particle motion. Both particles were slightly denser than water with specific gravity $\rho_p/\rho_f$ of 1.003 (denoted as P1) and 1.05 (denoted as P2). To investigate the effect of turbulent and coherent eddies on particle rotation, sphere diameters, $d^+$ of 60 and 120 wall units larger than the Kolmogorov length scale were studied. This corresponds to initial particle Reynolds numbers, $Re_p$ of 800 and 1900, with $Re_p$ defined as

$$Re_p = \frac{|U_p - \overline{U(y_p)}|d}{\nu} = \frac{U_{slip}\,d}{\nu}, \quad (1)$$

where $U_p$ is the particle streamwise translational velocity, $\overline{U(y_p)}$ is the *mean* streamwise fluid velocity at the height of the particle centroid, $U_{slip}$ is the local slip velocity, $d$ is the particle diameter and $\nu$ is the fluid kinematic viscosity. The particles were tracked over a streamwise distance up to $x = 6\delta$. For each case, 10 trajectories were captured using the same particle.

## 2.3 Reconstruction of particle translation and rotation

All image sequences were pre-processed before surface markers were tracked. First, the image of the sphere was isolated using a standard circular Hough Transform within MATLAB. A ring with thickness of four pixels was subtracted from the isolated sphere to avoid large pixel intensities exterior to the particle surface which could impact the tracking procedure by introducing *ghost* particles. The resulting sphere image was processed further by performing Gaussian smoothing with a 3 x 3 pixel window, and sharpening. Finally, pixel intensity values below an arbitrary intensity threshold were set to zero to avoid reconstruction of ghost or false particles. Fig. 3(a) presents an example of the original and pre-processed images from one camera in grayscale.

The optical system was calibrated using a two-level plate with round marks (LaVision Type 22) across nine planes in the spanwise direction spaced equally by 6.35 mm. The images acquired from all planes and cameras were used to detect circle patterns within the calibration routine of Davis 8.3. A third order polynomial fit was obtained for each plane and the obtained root-mean-square error of the grid point positions was 0.05-0.1 pixels. The pre-processed images from each camera pair were used to reconstruct the 3D coordinates of markers based on the mapping function of the volumetric calibration. The 3D reconstruction procedure, explained in detail by Wieneke (2008), was computed here using the 3D PTV routine in Davis 8.3. The markers were defined by the lowest intensity threshold, and their 2D centroid locations in each camera image were obtained using a Gaussian detection with fitting area of 3 x 3 pixels. A line of sight was generated for each marker from one camera using the mapping function. This line of sight was projected over the second camera image, and a triangulation error of 1 pixel was set to find the corresponding markers in the second camera image. Finally, 3D locations of reconstructed markers containing true and ghost markers originating from the ambiguities of intersected lines of sight from distinct 2D centroids were generated [Elsinga *et al.* (2011)]. An example of the reconstructed markers is illustrated in fig. 3(b).

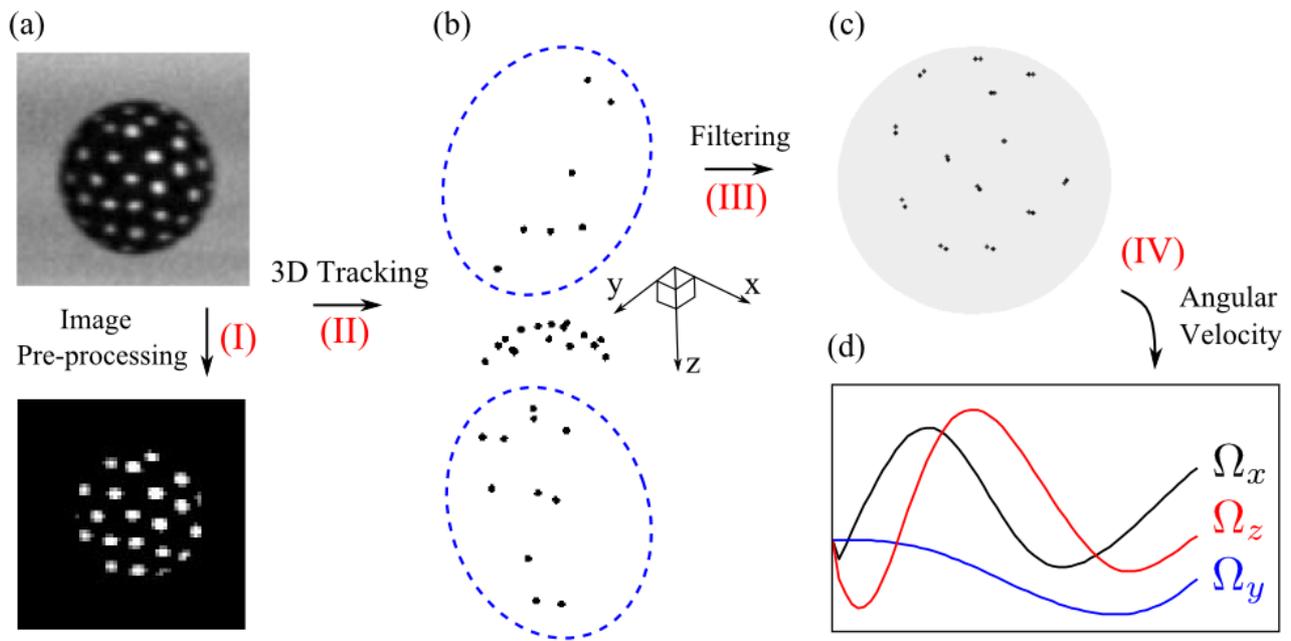

**Fig. 3** Steps to compute the translation and rotation of the sphere. (a) Example of image pre-processing. (b) 3D reconstructed markers including true and ghost (highlighted by dashed ellipses) particles. (c) Final retained tracked markers after filtering used to compute the sphere rotation by the optimal rotation matrix. (d) Sample result.

The data sets obtained from PTV were composed of the 3D coordinates of multiple markers (true and ghost markers) for each time step and their corresponding 3D velocity vectors. A filtering methodology was thus necessary to remove the ghost tracks to avoid bias errors. An example of the final retained tracked markers after filtering is illustrated in fig. 3(c). The center of the sphere was determined by the knowledge of at least three markers and the sphere radius: this allows us to obtain the translational dynamics of the sphere. These markers were further used to compute the sphere angular velocities from the optimal rotation matrix aligning both sets of points in consecutive images (see fig. 3(d)). The filtering methodology as well as the translation and rotation algorithms are explained in detail in Barros *et al.* (2018).

In general, the methodology relies on three particular aspects. First, the true markers are confined to a limited volume and depth. Thus, the median value of the depth coordinate ($z_{med}$) will tend to lie within this depth. A second key enabler to remove the ghost particles is the constraint over a hemisphere with known radius, which strictly enforces the quality of the retained particles within the tracking uncertainty. Finally, the use of time information to estimate $z_{med}$ and the center location from the previous time steps helps limit the search radius to find the true markers and the optimal sphere center coordinates. In this context, the translation results discussed below were computed by tracking the centroid of the sphere. The sphere images were pre-processed to become a circular white dot with radius of 5 pixels. After fitting the dot with a Gaussian smoothing filter, the dot was tracked in Davis 8.3 using the 3D-PTV routine. This computation is faster because only one particle is tracked in each time step and hence, this limits the generation of ghost particles. However, this method does not allow rotation computation. Therefore, the rotational dynamics of selected data sets were processed separately using the algorithm described earlier. Here, the 3D translation trajectories calculated from both methods agreed very well with each other.

The mean disparity error, $\epsilon_{disp}^*$ calculated by projecting the 3D reconstructed markers back to the camera image was 0.39 px. This gives an estimate of the uncertainty in the marker locations due to reconstruction errors (Wieneke (2008)). To reduce the noise in the displacement and rotation data for velocity and acceleration computation, the raw data was fitted to a quintic smoothing spline as introduced by Epps *et al.* (2010) before calculating the derivatives. Here, the uncertainty of the displacement and rotation data was computed based on the RMS between the raw and the smoothed data as proposed by Schneiders and Sciacchitano (2017). By considering all centroid coordinates, the mean displacement uncertainty was 0.75 px while the angular rotation uncertainty corresponds to 3 $rad/s$ which is equivalent to a maximum rotation displacement of 0.4 px per time step.

## 4. Results and Discussion

The wall-normal trajectories of spheres tracked over a streamwise distance of $5 - 6\delta$ are all plotted in fig. 4(a). For sphere P1 with $\rho_p/\rho_f = 1.003$, at both $Re_\tau$ values, the particle always lifts off when released. The sphere accelerates from rest with the fluid and ascends away from the wall. Meanwhile, at $Re_\tau = 700$, sphere P2 with $\rho_p/\rho_f = 1.05$ does not lift off but slides along the wall with the flow. At the higher $Re_\tau$ of 1300, as the mean shear becomes stronger, the denser sphere begins to lift off occasionally. The initial lift-off angles of the spheres fall between 2° and 10° with an average angle of 4.5° for P1 at $Re_\tau = 1300$ and 2.5° for the other two cases. As both spheres are slightly denser than water, they reach an initial peak height after the initial lift-offs before descending back towards the wall.

In each case, the sphere exhibits multiple translation behaviors. For sphere P1 at $Re_\tau = 1300$, four examples of trajectories are marked in fig. 5(a):

(i) Saltation (dashed line): after the initial lift-off, the sphere descends and collides with the wall. Then, it continues with a sequence of bouncing events.
(ii) Near-wall resuspension (dash-dotted line): the sphere lifts off at small magnitude and after descending towards the wall, the sphere reverses and ascends to a larger height without touching the wall.
(iii) Far-wall resuspension (solid line): the sphere lifts off to a larger magnitude and after a small descent, the sphere ascends again to a height farther from the wall.
(iv) Long projectile motion (dotted line): instead of an immediate lift-off, the sphere ascends gradually over the streamwise distance until reaching a maximum height before descending towards the wall.

For the same particle at lower $Re_\tau$ (see fig. 5(b)), saltation (v) and near-wall resuspension (vi) are also observed. Here, the sphere saltates more frequently at smaller magnitude. Considering both $Re_\tau$ values, the spheres resuspend up to a maximum $y^+$ of 120 and 280 wall units as $Re_\tau$ increases. These values are equivalent to $1.5d$ and $1.8d$ where $d$ is the sphere diameter.

For sphere P2 at $Re_\tau = 1300$, two common trajectories are observed (see fig. 5(c)). For the first trajectory (vii), the sphere always falls back and slides along the wall after initial lift-off. While sliding, sudden bouncing of varying magnitude is frequently observed. For the second trajectory (viii), no initial lift-off is observed. Instead, the sphere slides along the wall with random lift-offs. Unlike the saltation mode, the sphere propagation is a mix of sliding and lift-off events. At $Re_\tau = 700$, most of the runs exhibit similar behavior to trajectory (viii), and some of them slide entirely throughout the field of view without bouncing. Under this condition, the friction force plays an important role on the sliding and rolling motion along the wall.

As a given sphere lifts off and descends, it reaches a local maximum in height which we define here as a peak. If we compare the mean height of the initial lift-off peak for P1 as depicted in fig. 4(b), the sphere ascends further at $Re_\tau = 1300$ than at $Re_\tau = 700$ with an average $y^+$ of 133 and 40 wall units ($0.7d$ and $0.2d$), respectively. This shows that the height associated with the initial lift-off correlates strongly with the mean shear value. If we compare between particles P1 and P2 at $Re_\tau = 1300$, P2 ascends only to a mean height of 72 wall units ($0.1d$), which is around 14% of the P1 value. Although P2 is approximately 5% denser than P1, these observations indicate that the effect of the gravitational force on the initial lift-off of the sphere is significant. Hence, for an even denser sphere, a higher shear stress would be required to initiate the sphere lift-off.

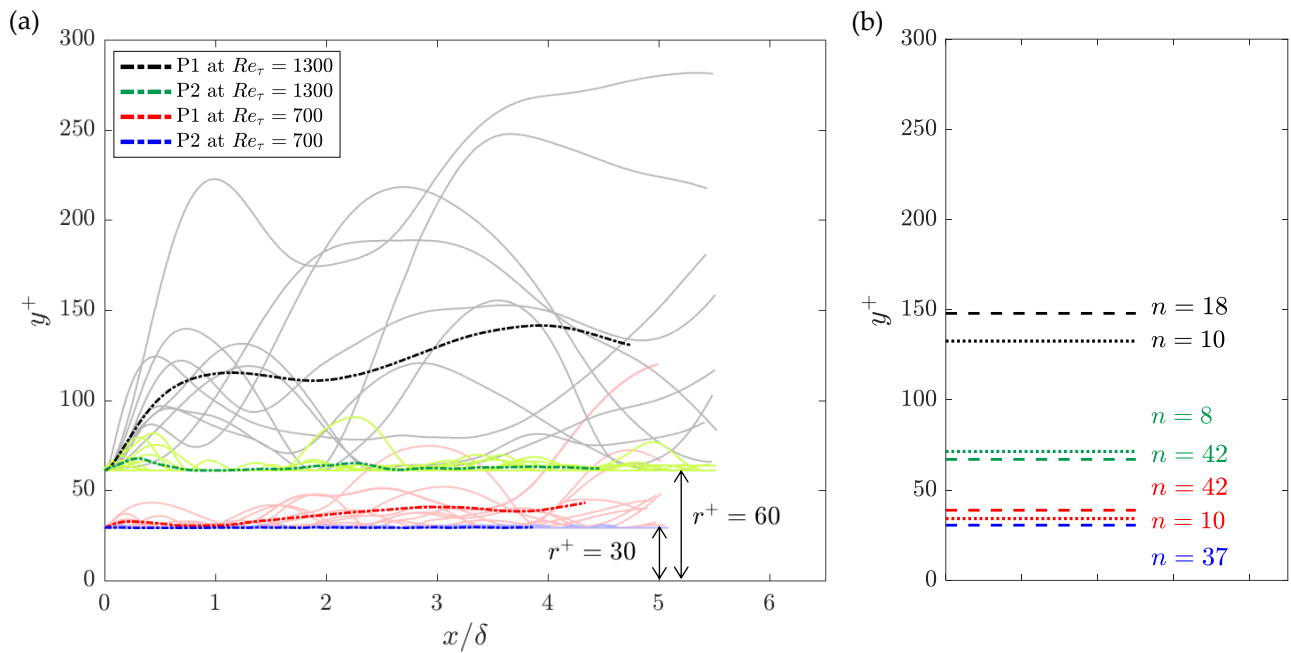

**Fig. 4** (a) Particle trajectories in wall-normal plane; Dashed lines represent the mean of 10 runs which are each depicted by a solid line. Specific gravity: P1 = 1.003 and P2 = 1.050. (b) Average particle wall-normal peak heights. Dashed lines represent $y^+$ values averaged across all peaks in 10 runs for each case while dotted lines represent the average height of the first $y^+$ peak for cases where the particle lifted off immediately after release. $n$ represents the number of peaks for each case.

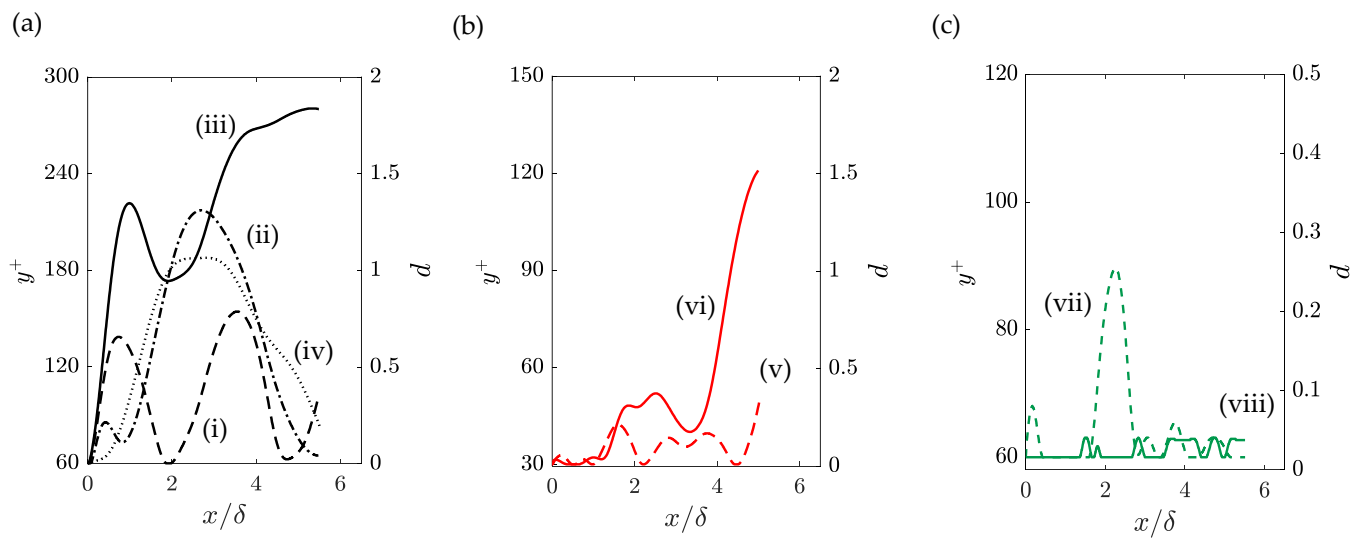

**Fig. 5** Examples of particle trajectories in wall-normal plane with left axis plotted in wall units, $y^+$ and right axis plotted in particle diameter, $d$. (a) P1 at $Re_\tau = 1300$, (b) P1 at $Re_\tau = 700$ and (c) P2 at $Re_\tau = 1300$.

Meanwhile, a comparison between the mean height of all peaks and the mean initial peak height for each case also shows some interesting features. For P1, the mean peak heights averaged over all peaks are higher than the mean initial peaks at both $Re_\tau$ values. This shows that on average, the sphere ascents extend farther from the wall after the initial lift-off ascent. The opposite result is observed for P2 at $Re_\tau = 1300$ where the mean peak height is smaller than the initial peak. As observed earlier, this sphere always descends back all the way to the wall after initial lift-off without resuspension. Although the sphere bounces randomly with varying magnitude during its sliding motion, the result indicates that the heights associated with these later bounces are mostly smaller than that reached after the initial lift-off.

The spanwise trajectories of the spheres are illustrated in fig. 6 as |z|, the magnitude of deviation from the origin. All trajectories are distributed symmetrically about z=0 with some crossings at z=0. Even though the particle wall-normal behaviors are different in each case, the spheres migrate with mean and maximum spreading angles of approximately ±4° and ±7° in all cases. In this context, the spheres migrate up to $\pm 0.6\delta$ or $7d$ in the spanwise direction, which is more than three times the maximum wall-normal lift-off observed.

Now that we have presented the multiple features of particle trajectories, we turn our attention to the particle translational velocities. Based on the particle streamwise velocity in fig. 7(a), all spheres accelerate with a steep slope once released and eventually approach an approximate terminal velocity. After the initial acceleration, fluctuations of order $0.2 U_\infty$ are observed in many individual runs. The mean streamwise velocities show that as $Re_\tau$ increases, the particle translational velocity increases, while P1 translates faster than P2 at both $Re_\tau$. In other words, the denser sphere P2 at $Re_\tau = 700$ propagates most slowly among all cases. As this sphere slides along the wall, friction opposes and retards its forward motion.

Using the particle wall-normal location, the mean fluid velocity at the sphere centroid is computed as illustrated in fig. 7(b). As P1 ascends to greater heights in the boundary layer, the mean fluid velocity also increases. In this context, the particle gains more momentum from the fluid to accelerate. Meanwhile, for sphere P2, the fluid velocity along the trajectories varies minimally with maximum increment of $0.03 U_\infty$ as compared to $0.15 U_\infty$ for P1 at $Re_\tau = 1300$.

Fig. 7(c) depicts the particle Reynolds number computed based on the slip velocity defined in equation (1). In all cases, there is a strong acceleration phase immediately after the release of the particle corresponding with a strong decrease in $Re_p$. Here, the lift force acting on a sphere is defined as $F_L = \frac{\pi}{8}\rho_f U_{slip}^2 C_L d^2$ where $C_L$ is the lift coefficient. At the instant the sphere is released at $Re_\tau = 1300$, it experiences a stronger lift force due to the larger slip velocities. Hence, a sphere with higher initial $Re_p$ lifts off to greater heights than one with lower initial $Re_p$. Also, as the gravitational force opposes the sphere's upward motion, the initial lift-off height of sphere P2 is always smaller than that for sphere P1 under the same flow condition.

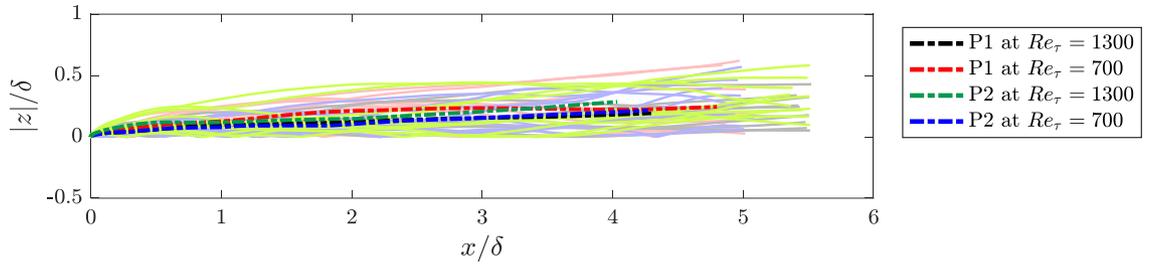

**Fig. 6** Particle absolute spanwise trajectories, $|z|$.

As the particles translate up to $x = 1\delta$, the $Re_p$ for all runs decreases sharply due to the reduction in the particle slip velocities. As the slip velocity decreases, the lift force acting on the sphere must also decrease assuming $C_L$ is constant. This reduction is stronger for $Re_\tau = 1300$ as $Re_p$ decreases by almost an order of magnitude for sphere P1. Any shear-induced lift-off events should be greatly suppressed after $x = 1\delta$. However, as described earlier, the particle continues to resuspend to greater heights or bounce frequently beyond this location. This suggests that in addition to any lift force due to shearing, there exists an additional upward mechanism whose strength varies with time. As the upward and downward motions are occurring in the logarithmic region, we hypothesize that the particles are dragged upward and downward by hairpins, or other coherent flow structures present there. This phenomenon has been visualized by van Hout (2013) through PIV and PTV experiments where the lift-off behavior of the sphere was correlated with the size and strength of the coherent structures encountered.

Particle wall-normal ($V_p$) and spanwise ($W_p$) velocities are plotted in fig. 8(a) and (b), respectively. Sphere P1 at $Re_\tau = 1300$ has the highest $V_p$ values among all cases and hence it tends to move further away from the wall. By contrast, sphere P2 at $Re_\tau = 700$, which generally bounces close to the wall, has the smallest $V_P$ values and a mean $V_P$ centered near zero. Based on the mean spanwise velocities, spheres that initially lift off from the wall have larger initial spanwise velocities. In all cases, the spanwise velocities fluctuate over a range of $\pm 0.1 U_\infty$ which is clearly larger than the $\pm 0.05 U_\infty$ range observed for wall-normal velocity. The magnitude of spanwise migration is thus larger than the wall-normal lift-off.

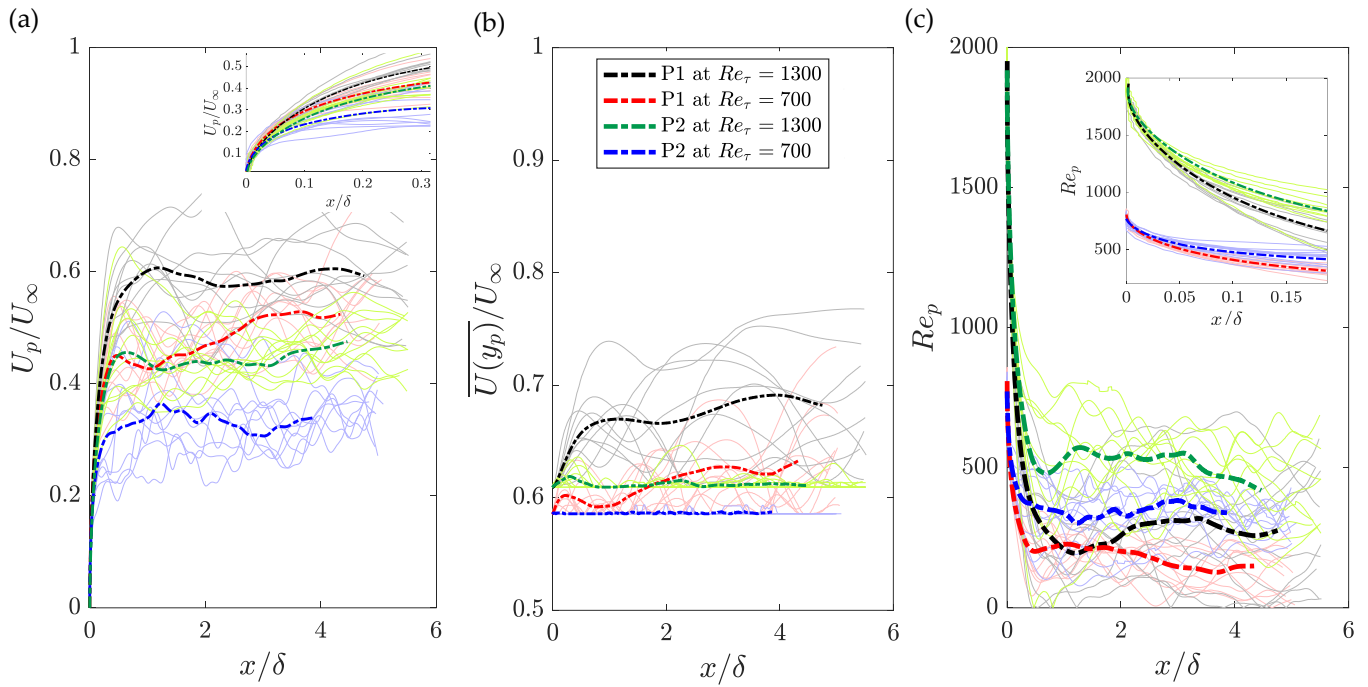

**Fig. 7** a) Particle streamwise velocities, $U_p$ and b) mean fluid velocity based on particle wall-normal location, $\overline{U(y_p)}$, both normalized by the streamwise free-stream velocity, $U_\infty$ at respective $Re_\tau$. c) Particle Reynolds number calculated based on equation (1).

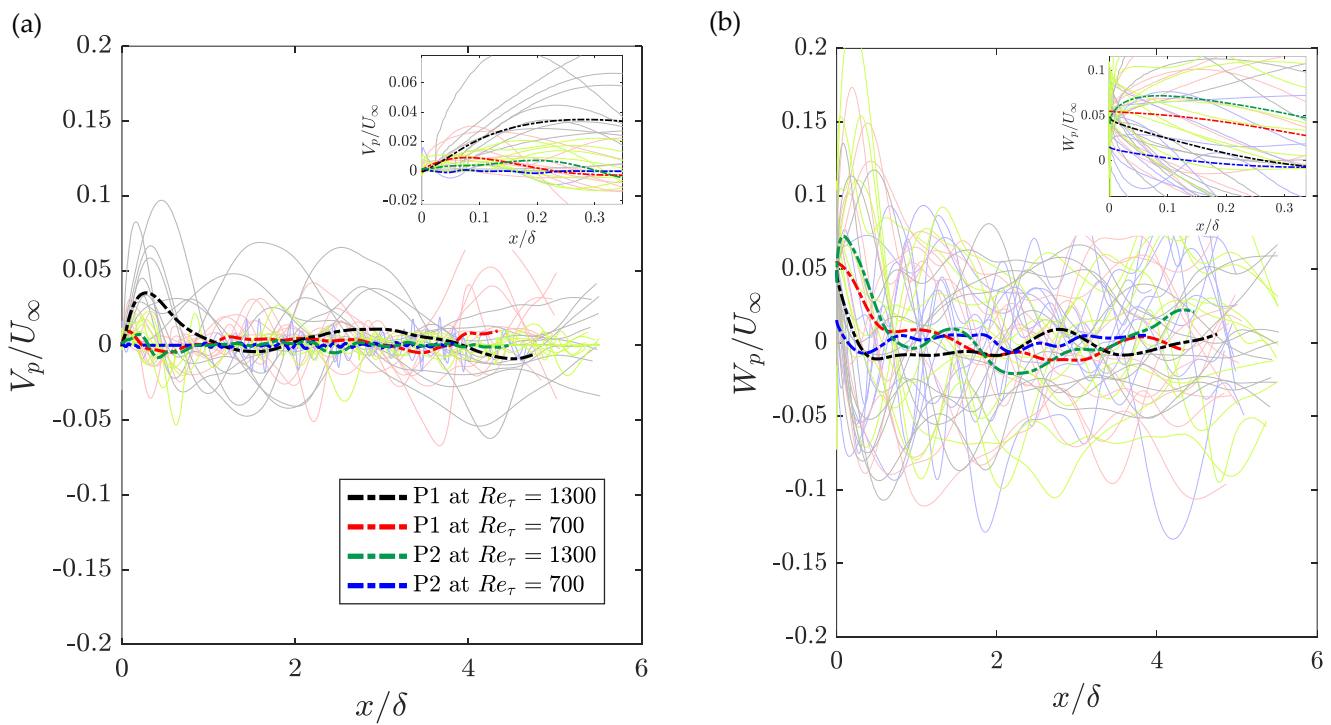

**Fig. 8** Particle a) wall-normal velocities, $V_p$ and b) spanwise velocities, $W_p$ both normalized by the streamwise free-stream velocity, $U_\infty$ at respective $Re_\tau$.

In fig. 9, we present the three components of the angular velocity, $\Omega$ for a set of particle trajectories at $Re_\tau = 1300$. The corresponding wall-normal trajectories are also depicted in the plots. Generally, the measurements indicate the presence of sphere rotations about all three axes. The magnitude of the normalized spinning rate $\Omega^* = \Omega d/2U_\infty$ lies in the range [-0.05, 0.05], with few exceptions exceeding [-0.075, 0.075]. This range corresponds to 5% of the free-stream velocity of the boundary-layer, representing relatively small rotational velocities. While rotation about $x$ and $y$ might appear with both signs, we note that spanwise spinning ($z$) is predominantly negative. In principle, this is expected due to the negative spanwise vorticity associated with the boundary-layer profile.

The spanwise rotation presents systematic quasi-periodic oscillations for both particles. Sphere P1 initially rotates with $\Omega_z^*$ up to -0.05 and this angular velocity is gradually damped as the sphere moves forward (see fig. 9(a)). Although each of the wall-normal trajectories exhibit quite distinct behaviors, they do not appear to impact the magnitude of $\Omega_z^*$ significantly. Also, the wavelength of $\Omega_z^*$ fluctuations does not correlate with the measured variations of particle height $y^+$. For the denser particle (P2), the behavior of $\Omega_z^*$ differs: while $\Omega_z^*$ oscillates, the wavelength of such fluctuations over $x$ is considerably shorter, both for sliding and bouncing trajectories. Interestingly, a bouncing event in one of the trajectories is responsible for the sudden increase in $\Omega_z^*$ to a value close to -0.15. This angular acceleration is unambiguously the result of the additional torque exerted on the sphere by the frictional force upon wall contact.

Although the mean shear in the boundary layer might presumably induce a rotation about $z$, we also note significant rotations about $x$ and $y$. Fig. 9(b) shows that particle P1 displays low-frequency rotational dynamics about the wall-normal axis in the range [-0.04, 0.04]. Meanwhile, the values of $\Omega_y^*$ for the denser particle are generally smaller, although a significant acceleration about $y$ appears following the bouncing event mentioned above. For the lighter particle, the rotation about the streamwise axis ($\Omega_x^*$) presents similar fluctuations as those of $\Omega_z^*$. It is additionally damped further downstream as shown in fig. 9(c). On the other hand, the values of $\Omega_x^*$ in the trajectories of particle P2 show distinct behaviors, which might be the result of the complex interaction between the bouncing or sliding particle and the wall.

The reduction of $\Omega_x^*$ and $\Omega_z^*$ measured further downstream for particle P1 was evidenced qualitatively by Francis (1973), who also concluded that spherical particles rotate less than other grain geometries. It is of interest to further investigate the origins of angular momentum exchange between the fluid flow and the sphere. The mean shear undoubtedly plays a role. The convection of near-wall coherent structures might also play a significant role on the angular dynamics of the sphere. Another mechanism that deserves further investigation is the impact of the sphere wake at these values of $Re_P$, which can potentially generate rotation about $x$ and $y$ due to the random asymmetries of the wake.

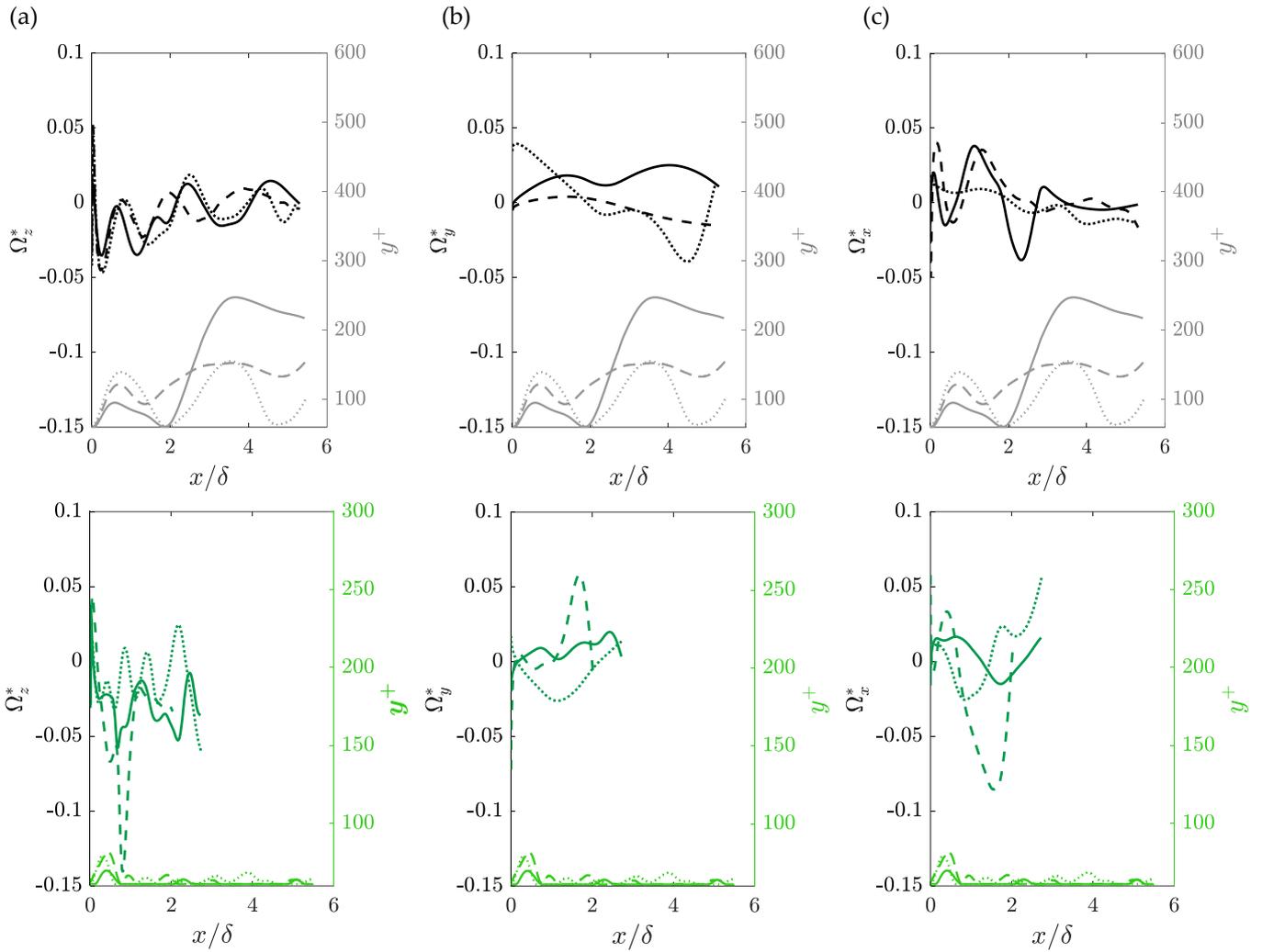

**Fig. 9** Particle normalized spinning rate $\Omega^*$ about (a) spanwise $z$, (b) wall-normal $y$ and (c) streamwise x - axes (left axes) and the corresponding wall-normal trajectories (right axes) at $Re_\tau = 1300$. Top figures (black): sphere P1; Bottom figures (green): sphere P2.

## 5. Conclusions

The 3D translational dynamics of two nearly neutrally buoyant spheres are investigated at $Re_\tau$ of 700 and 1300. Based on the wall-normal trajectories, we observe that the particle lift-off activities are limited to the logarithmic layer. Once released, sphere P1 with $\rho_p/\rho_f = 1.003$ always lifts off from the wall to a greater magnitude at $Re_\tau = 1300$ than $Re_\tau = 700$, which we assume is due to the stronger mean shear. Meanwhile, as sphere P2 is more dense ($\rho_p/\rho_f = 1.050$), initial lift-off is observed only at $Re_\tau = 1300$ where the mean shear is higher. After the initial lift-off, both spheres ascend to an initial peak height before descending towards the wall due to gravity. At $Re_\tau = 1300$, sphere P1 exhibits multiple behaviors including saltation, near-wall resuspension, far-wall resuspension and projectile motion over a relatively long distance. As $Re_\tau$ decreases, this sphere propagates with either saltation or near-wall resuspension with more frequent lift-off activities and smaller peak heights. By contrast, the denser sphere at higher $Re_\tau$ tends to fall back to the wall after the initial lift-off and propagate with a combination of sliding and bouncing events. Although this sphere does not lift-off at lower $Re_\tau$, small occasional bounces are observed in some runs. These observations show that for a sphere propagating under a given flow condition, both the translational behavior and lift-off magnitude can vary significantly. Generally, the lift-off magnitudes for P1 and P2 fall within $2d$ and $0.5d$, respectively, with the larger magnitude at $Re_\tau = 1300$. Also, a comparison between P1 and P2 highlights that even a 5% increase in density can have a significant effect on the sphere translational dynamics.

After the strong initial streamwise particle acceleration up to $x = 1\delta$, the mean shear induced lift force must decrease sharply. An additional lift-off mechanism is thus required to generate the subsequent lift-off events and upward motions observed frequently in the various cases. Based on the results by van Hout (2013), we hypothesize that these lift-off events are initiated and driven by coherent flow structures in the logarithmic region. Here, the sphere translational motions described above would vary depending on the size and strength of the structures encountered. Although the spheres exhibit different wall-normal behaviors in each case, the spheres migrate in the spanwise direction with mean and maximum spreading angles of around $\pm 4°$ and $\pm 7°$ in all cases.

The rotational dynamics of both spheres at $Re_\tau = 1300$ are investigated based on their angular velocities. The results indicate the presence of rotations about all three axes with normalized velocities of approximately 5% of $U_\infty$. Based on the selected runs studied, both spheres always rotate about the spanwise axis in a quasi-periodic manner. Similar behaviors are observed in the streamwise rotations of the lighter sphere but more random oscillations are observed in both streamwise and wall-normal rotations of the denser sphere. Aside from the mean shear which may initiate spanwise rotations, these complicated dynamics could be due to the perturbations from coherent structures in the surrounding flow as well as wake structures associated with the spheres. In all cases where the sphere propagates along the wall, sliding motion, rather than forward rolling motion, is dominant.

**Acknowledgments**

The authors are grateful to Nicholas Morse and Ben Hiltbrand for their efforts in fabricating the magnetic wax particle and developing the particle rotation algorithm. This work is funded by National Science Foundation (NSF) under CBET-1510154.